\documentclass[prb,superscriptaddress, twocolumn]{revtex4-1}
\usepackage{xcolor}
\usepackage{graphicx}
\usepackage{amsmath}
\usepackage[normalem]{ulem}

\begin{document}
\title{
An accurate tight binding model for twisted bilayer graphene \\ 
describes topological flat bands without geometric relaxation
}
\author{Shivesh Pathak}
\affiliation{Department of Physics, Institute for Condensed Matter Theory, National Center for Superconducting Applications, University of Illinois at Urbana-Champaign}
\author{Tawfiqur Rakib}
\affiliation{Department of Mechanical Science and Engineering, Department of Materials Science and Engineering, and Materials Research Laboratory, University of Illinois at Urbana-Champaign}
\author{Run Hou}
\affiliation{Department of Physics and Astronomy,  Rice University}
\author{Andriy Nevidomskyy}
\affiliation{Department of Physics and Astronomy,  Rice University}
\author{Elif Ertekin}
\affiliation{Department of Mechanical Science and Engineering, Department of Materials Science and Engineering, and Materials Research Laboratory, University of Illinois at Urbana-Champaign}
\author{Harley T. Johnson}
\affiliation{Department of Mechanical Science and Engineering, Department of Materials Science and Engineering, and Materials Research Laboratory, University of Illinois at Urbana-Champaign}
\author{Lucas K. Wagner} 
\affiliation{Department of Physics, Institute for Condensed Matter Theory, National Center for Superconducting Applications, University of Illinois at Urbana-Champaign}

\begin{abstract}
A major hurdle in understanding the phase diagram of twisted bilayer graphene (TBLG) are the roles of lattice relaxation and electronic structure on isolated band flattening near magic twist angles.
In this work, the authors develop an accurate local environment tight binding model (LETB) fit to tight binding parameters computed from \textit{ab initio} density functional theory (DFT) calculations across many atomic configurations.
With the accurate parameterization, it is found that the magic angle shifts to slightly lower angles than often quoted, from around 1.05$^\circ$ to around 0.99$^\circ$, and that isolated flat bands appear for rigidly rotated graphene layers, with enhancement of the flat bands when the layers are allowed to distort. 
Study of the orbital localization supports the emergence of fragile topology in the isolated flat bands without the need for lattice relaxation.
\end{abstract}
\maketitle

\section{Introduction}
An exciting new task in condensed matter physics is understanding the microscopic mechanisms leading to a diverse set of electronic states in twisted bilayer graphene (TBLG).
TBLG is a van der Waals structure that consists of two sheets of single layer graphene laid on top of each other and twisted with a relative twist angle $\theta$.
For certain ``magic'' values of $\theta$, the first of which has been cited as near  1.05$^\circ$,\cite{cao1} TBLG hosts correlated insulating and superconducting  phases\cite{cao2, cao1, PhysRevLett.117.116804,independent,delocalized} and anomalous Hall effects\cite{orbital, symmetry} 
It is believed that these correlated phases emerge through the interplay of electronic and structural degrees of freedom\cite{PhysRevB.93.235153, PhysRevResearch.2.043127, PhysRevB.96.075311}.

The rich phase diagram of TBLG has been suggested to arise from band flattening at the Fermi level\cite{cao1, cao2}.
The current understanding is that with sufficiently flat bands, even weak effective interactions can dominate the low-energy behavior, leading to the observed correlated states in TBLG near magic twist angles.
The bulk of evidence for band flattening in TBLG is found in theoretical calculations\cite{PhysRevB.96.075311, PhysRevB.86.125413, PhysRevB.82.121407, Bistritzer12233, PhysRevB.100.205114, PhysRevLett.122.106405} where bands near the Fermi level are shown to flatten to a few meV bandwidth near magic twist angles.
Experimentally,  a recent combination of low-energy electron microscopy and angle-resolved photoemission spectroscopy measurements\cite{flatbands} also observe flat bands at charge neutrality with bandwidths of 30 $\pm$ 15 meV at $\theta = 1.3^\circ$.

Structural relaxations accompany the changes in electronic structure near the magic twist angles.
In TBLG, structural relaxations enhance the size of low-energy AB regions and constrict those of AA regions and are paired with out-of-plane buckling, bringing together AB regions and pushing apart AA regions.
This structural relaxation has been observed experimentally through scanning tunneling microscopy measurements on TBLG\cite{Jiang2019} and in theoretical calculations
\cite{PhysRevB.96.075311, fang2019angledependent, PhysRevB.90.155451, doi:10.1021/acs.nanolett.6b02870, Jain_2016, PhysRevResearch.2.043127, PhysRevResearch.1.013001}.
Using simple tight binding models --- such as the one of Moon and Koshino (MK)\cite{PhysRevB.85.195458} --- it has been proposed that lattice relaxation is required to maintain the fragile topology\cite{PhysRevLett.121.126402, PhysRevLett.121.106403, PhysRevB.99.195455, PhysRevLett.123.036401, PhysRevB.98.085435, PhysRevX.9.021013,PhysRevLett.124.167002} of the flat bands and their energetic isolation\cite{PhysRevB.96.075311} from the rest of the bands in the system.

Thus it appears that the structural and electronic degrees of freedom are tightly coupled in this system.
However, the conclusion that lattice relaxation is required for energetic isolation and fragile topology of flat bands --- indicators of tightly coupled electronic and structural degrees of freedom --- were derived using phenomenological tight binding models which have no direct link to \textit{ab initio} simulation.
As such, it remains to be seen whether accurate, \textit{ab initio}, treatment of the electronic and structural degrees of freedom would arrive at the same conclusions.
While some density functional theory (DFT) calculations have been performed at the magic angle scale,\cite{PhysRevB.86.125413, PhysRevB.90.155451} these calculations are very computationally demanding and it is not feasible to perform many calculations to disentangle electronic and structural degrees of freedom.

In this manuscript, we present a highly accurate local environment tight-binding (LETB) model for TBLG, fit to  DFT calculations of 72 structural configurations of bilayer graphene.
We show that the LETB model reproduces DFT for structural configurations relevant to TBLG much more accurately than simpler TB models.
In the LETB model,  isolated flat bands with fragile topology are observed both with and without lattice relaxation, in contrast to the current understanding in which lattice relaxation is required for isolated flat bands with fragile topology.
A Python package that generates LETB models for any atomic configuration is made available\cite{letb}.

\section{Training data for twisted bilayer graphene model}
\subsection{Atomic configurations}
Our goal is to develop a TB model that correctly accounts for the variations in structure of TBLG.
Our approach is to use deformed, untwisted, primitive cell bilayer graphene configurations.
We use two deformation strategies in order to capture the variations in stacking pattern as well as in-plane and out-of-plane relaxations seen in small twist angle TBLG. 
The first are in-plane and out-of-plane shifts, strains and shears, and the second are random atomic variations.
Combined we consider 72 different deformed configurations of bilayer graphene.
Details of the deformation strategies follow.

We begin by introducing notation for the deformed atomic configurations.
There are four atoms in our configurations with positions denoted by the vectors
$\vec{a}_1, \vec{b}_1, \vec{a}_2, \vec{b}_2,$
where $a, b$ refer to the two distinct atoms in a given graphene layer, and $1, 2$ are layer indices.
The real space lattice vectors for the primitive cell are denoted by $\vec{L}_1, \vec{L}_2$.
These six vectors fully describe the atomic configurations.
 
The shifted, strained, and sheared configurations are best understood through the deformation equation
\begin{equation}
\begin{split}
\begin{bmatrix}
\vec{a}_1 \\
\vec{b}_1 \\ 
\vec{a}_2 \\ 
\vec{b}_2 \\
\vec{L}_1 \\
 \vec{L}_1 \\
\end{bmatrix}
= 
\left(
\begin{bmatrix}
0 & 0 & 0\\
0 & a_0 & 0\\
0 & 0 & c_0 \\
0 & a_0 & c_0 \\
\sqrt{3} a_0& 0& 0\\
-\frac{\sqrt{3}}{2} a_0 & \frac{3}{2} a_0 & 0\\
\end{bmatrix} 
+ 
\begin{bmatrix}
0 & 0 & 0\\
0 & 0 & 0\\
0 & s & \Delta \\
0 & s & \Delta \\
0 & 0 & 0\\
0 &0 & 0\\
\end{bmatrix} 
\right) 
\times
\\
\begin{bmatrix}
1 + \epsilon_{xx} & \epsilon_{xy}\\
\epsilon_{xy} & 1 + \epsilon_{yy}\\
0 &0\\
\end{bmatrix}
\end{split}
\label{eq:sss_deform}
\end{equation}

The first matrix on the right hand side denotes our reference configuration: the equilibrium AA configuration with lattice constant $a_0 = 2.683$ Bohr and interlayer separation  $c_0 = 6.646$ Bohr.
The second matrix applies the relative shift between the two layers  through the parameter $s$ and inter-layer spacing variations through $\Delta$. 
In our dataset, $s$ takes three values - $0$, $a$, $3a/2$ - which correspond to the AA, AB and SP bilayer configurations, and $\Delta$ takes three values - 0 Bohr, -0.149 Bohr, -0.126 Bohr - corresponding to the AA, AB, and SP layer spacings.
The third matrix applies the in-plane shears via $\epsilon_{xy}$ and in-plane strains in the x- and y- directions via $\epsilon_{xx}, \epsilon_{yy}$.
All three parameters independently take three values, -0.01, 0, and 0.01, corresponding to 1\% atomic position variations in-plane.
The five parameters, $s, \Delta, \epsilon_{xx}, \epsilon_{xy}, \epsilon_{yy}$ can be varied together, leading to a large space of strained, sheared, and shifted configurations of the bilayer graphene.

For the random configurations, the parameters $s, \Delta, \epsilon_{xx}, \epsilon_{xy}, \epsilon_{yy}$ are chosen at random from a uniform distribution between the corresponding ranges described in the previous paragraph.
Once this new deformed configuration is constructed, we add an additional $6 \times 3$ random matrix $R$ to the configuration to allow for arbitrary inter-atomic displacements not captured by Eq.~\eqref{eq:sss_deform}.
To ensure that the random displacements incurred by $R$ are not too large, we require that the Frobenius norm $||R||_F = 0.01a_0$, corresponding to random variations that move the atomic configurations by a percent of the equilibrium atomic spacing.
We reserve two twisted configurations at 9.4$^{\circ}$ and 4.4$^{\circ}$ to test the models.

\subsection{Tight binding parameters from density functional theory (DFT)}
\begin{figure*}
\centering
\includegraphics[scale=0.50]{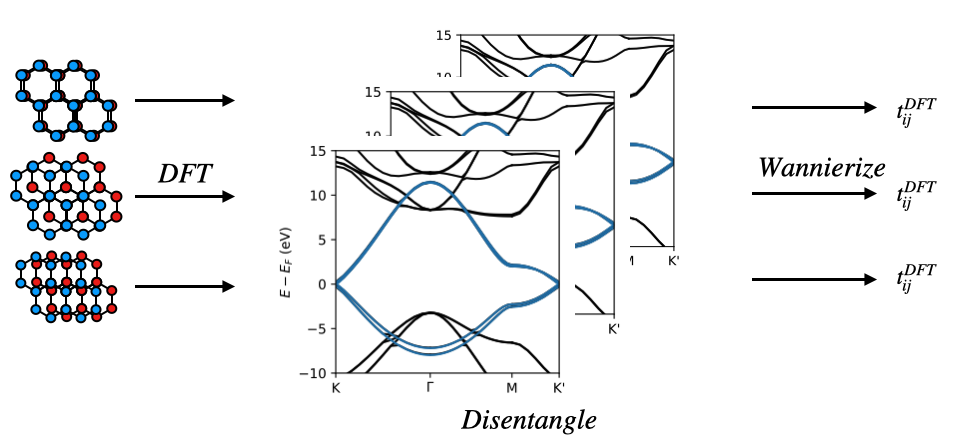}
\caption{Workflow for computing tight binding parameters from atomic configurations.
First, density functional theory is used to compute the band structure.
Next, the low-energy $\pi$ bands are disentangled, and these bands are Wannierized to return the first-principles tight binding parameters.
}
\label{fig:workflow}
\end{figure*}

In Figure~\ref{fig:workflow} we illustrate our workflow for computing the training tight binding parameters from atomic configurations.
First, we compute band structures for each configuration using van der Waals DFT.
Then we extract the $\pi$ bands near the Fermi level and Wannierize these low-energy bands.
The Wannierization procedure returns tight binding parameters for the low-energy bands which we want to model. 
Details of the procedure follow.

For each of the deformed geometries in our training data set, we computed the total SCF energies and band structure using van  der Waals DFT.
We used the BEEF-VDW van der Waals functional\cite{PhysRevB.85.235149}, a polarized triple-zeta all electron basis constructed for solid-state DFT calculations\cite{basis}, and a 36 $\times$ 36 $\times$ 1 k-point grid.
All DFT calculations were carried out using the PySCF package\cite{https://doi.org/10.1002/wcms.1340, doi:10.1063/5.0006074}.
To demonstrate the accuracy of the DFT calculations, we present a comparison of the DFT band structure to angle resolved photoemission spectroscopy (ARPES) in the Appendix.

From each DFT calculation we isolate the four $\pi$ bands and Wannierize them to extract tight binding (TB) parameters.
The disentanglement scheme of Souza, Marzari, and Vanderbilt\cite{PhysRevB.65.035109} is used to isolate the four $\pi$ bands from bands with $C\ 1s,\ 2s,\  2p_x, \ 2p_y$ orbital character.
The maximally localized Wannierization procedure\cite{PhysRevB.56.12847, RevModPhys.84.1419, doi:https://doi.org/10.1002/9781119148739.ch6} is then used to generate maximally localized Wannier orbitals (MLWO), and the isolated $\pi$ bands are projected onto the MLWOs to obtain TB parameters.
All Wannierization calculations are carried out using the Wannier90\cite{MOSTOFI20142309} and pyWannier90 packages\cite{doi:10.1063/5.0006074}.
A comparison of the Wannierized band structure to DFT is presented in the Appendix.

\subsection{Combined dataset}
The full dataset containing deformed bilayer atomic configurations and the corresponding tight binding parameters computed from DFT is available online alongside the model Python package\cite{letb}.
The dataset contains 72 different data files, one for each atomic configuration as described in section II A.
Each data file has three pieces of information: the atomic configuration, the DFT total energy, and the tight binding parameters from DFT.
The atomic configurations are stored as a list of three lattice vectors and a list of four atomic basis vectors in units of Angstrom.
The tight binding parameters are stored via five different lists encoding the hopping magnitude between two atoms.
The first pair of lists labeled \texttt{atomi, atomj} indicate which two atoms the hopping occurs between, relative to the atomic basis list.
The next pair labeled \texttt{displacementi, displacementj} indicate how many lattice vector displacements in the $x, y$ direction are between the hopping centers.
The last of the five lists,  \texttt{tij} is the hopping value in Hartree.

\section{Local environment tight binding model (LETB)}
We propose a local environment dependent TB parameterization which in addition to the interatomic separation, explicitly accounts for the detailed nuclear configuration in the vicinity of atoms involved in hopping.
The general form of the LETB is
\begin{equation}
H_{LETB} = \sum_{ij\sigma} t^{LETB}_{ij}(\vec{R}_i, \vec{R}_j, \{\vec{R}_{ij}\}) c_{i\sigma}^\dagger c_{j\sigma} + h.c. 
\end{equation}
Here $i, j$ are atomic indices for the pair of atoms with a hopping value of $t_{ij}^{LETB}$, $\sigma$ is the spin index, $c_{i\sigma}^\dagger$ is the creation operator for a localized orbital on site $i$ of $p_z$ character with spin $\sigma$;  $\vec{R}_i,$ is the location of atom $i$, and $\{\vec{R}_{ij}\}$ is a set of nuclear positions within the local environment of atoms $i,  j$.
The definition of the local environment of two atoms $i, j$ and functional dependence of $t_{ij}$ on the local environment coordinates are made explicit in sections III A and III B. 

The LETB can be contrasted with the MK model\cite{PhysRevB.85.195458}: 
\begin{equation}
\begin{split}
H_{MK} = \sum_{ij\sigma}t^{MK}_{ij}(\vec{R}_i - \vec{R}_j)c_{i\sigma}^\dagger c_{j\sigma} + h.c. \\
t^{MK}_{ij}(\vec{d}) = 
V_{pp\pi}^0 e^{-(\frac{|\vec{d}| - a_0}{\delta})}
\Big[1 - (\frac{\vec{d} \cdot \hat{z}}{|\vec{d}|})^2\Big] \\ 
V_{pp\sigma}^0 e^{-(\frac{|\vec{d}| - d_0}{\delta})}(\frac{\vec{d} \cdot \hat{z}}{|\vec{d}|})^2
\end{split}
\label{eq:MK}
\end{equation}
where the constants take values $V_{pp\pi}^0 = -2.7 $eV,   $V_{pp\sigma}^0 = 0.48$ eV,  $a_0 = 2.683$ Bohr,  $d_0 = 6.331$ Bohr,  $\delta = 0.246 a_0.$
Conceptually, this model bridges two exponentials: the $V_{pp\pi}$ exponential which corresponds to intra-layer matrix elements and the $V_{pp\sigma}$ term corresponding to inter-layer matrix elements.
Unlike the present LETB model, the MK hoppings depend only on the pairwise displacement vector between hopping centers.
THe MK parameterization is commonly used for TBLG band structure calculations, and we use it as a point of reference.

\subsection{Intra-layer hopping}
\begin{figure}
\centering
\includegraphics{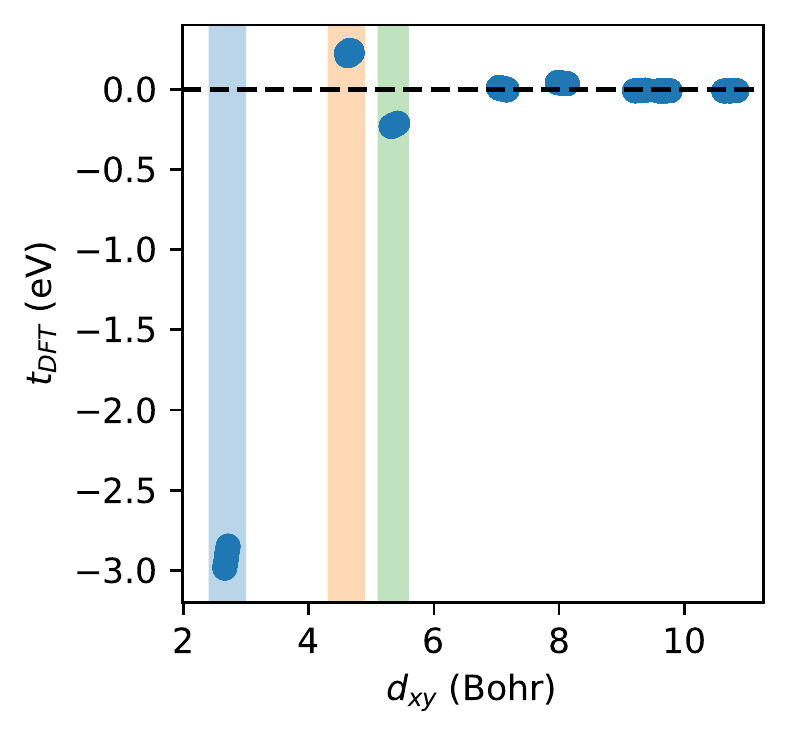}
\caption{Intra-layer hoppings in sampled configurations as a function of the in-plane distance $d_{xy}$, computed from the DFT calculations. 
The blue, orange, and green regions denote the first, second, and third nearest neighbor regions defined by $\pm $5\% intervals around the the equilibrium distances of $a_0\ , \sqrt{3}a_0$ and $2a_0$ respectively.
A clear separation of nearest neighbors is present, allowing for separate models to be fit for each term.
}
\label{fig:intra-layer-d}
\end{figure}

Fig~\ref{fig:intra-layer-d} shows the intra-layer hopping obtained by Wannierizing the DFT bands as a function of the in-plane distance $d_{xy}$, which is the distance projected onto the $xy$ plane.
The intra-layer hopping decreases with the in-plane distance very quickly, to the order of 0.01 eV by the fourth nearest neighbor. 
We thus focus on obtaining an accurate model up to the third nearest neighbor. 
This choice will be later justified by a low error in computed band structures using the local TB model, compared to the reference DFT band structures. 

\begin{figure*}
\centering
\includegraphics[scale=0.45]{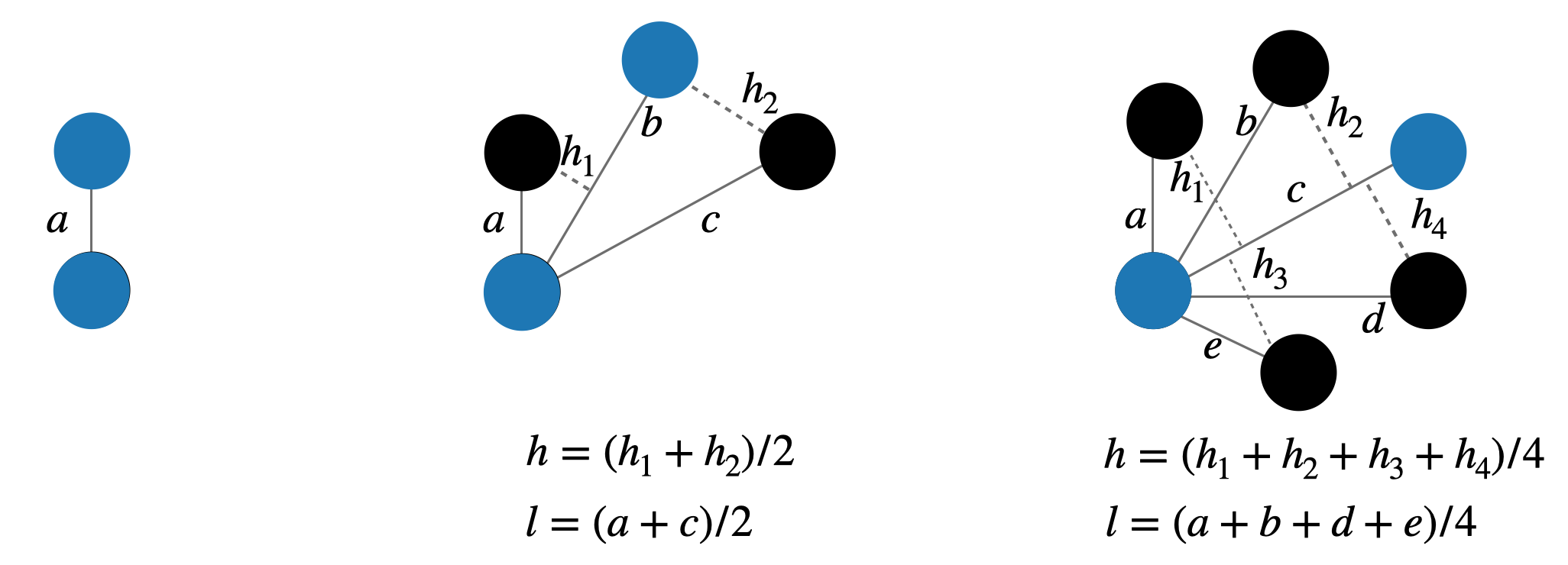}
\caption{Schematic diagram listing all possible descriptors used to fit linear models for intra-layer hopping.
The blue atoms denote the atomic pairs between - from left to right - nearest, next nearest, and third nearest neighbor atoms.
Solid lines denote direct distances between atoms and dashed lines perpendicular distances. 
Composite descriptors also considered in regression are listed below the schematics.
}
\label{fig:descriptors}
\end{figure*}

\begin{table}
\centering
\caption{LETB cross validated fit parameters. 
The reported values with parenthetical errors are the mean and standard deviation of the parameters over five folds.
The Fang parameters were refit to our DFT data.}
Intra-layer Hopping Models \\
\begin{tabular}{|c|l|}
\hline
\ 1-NN \ & \ -9.68(4) + 2.52(1)$a$                                \\
\ 2-NN \ & \ 1.55(1) + 0.022(2)$b$ - 0.66(1)$h_1$ - 0.20(1)$h_2$ \ \\
\ 3-NN \ & \ -1.23(1) + 0.04(1)$c$ - 0.12(1)$h$ + 0.23(1)$l$   \\
\hline
\end{tabular}
\linebreak 

Fang \cite{fang2019angledependent} Inter-layer Hopping Model \\
\begin{tabular}{|c|c|c|c|c|}
\hline
$i\ \ \ $ & $\lambda_i$ (meV) & $\xi_i$ & $x_i$   & $\kappa_i$ \\
\hline
0   & 239(2)            & 2.12(2) &         & 1.871(4)   \\
3   & -40(1)            & 3.8(4)  & 0.52(4) &            \\
6   & -5.9(7)           & 6.0(8)  & 1.52(1) & 1.73(2)    \\
\hline
\end{tabular}

\label{table:parameters}
\end{table}

\begin{figure}
\centering
\includegraphics{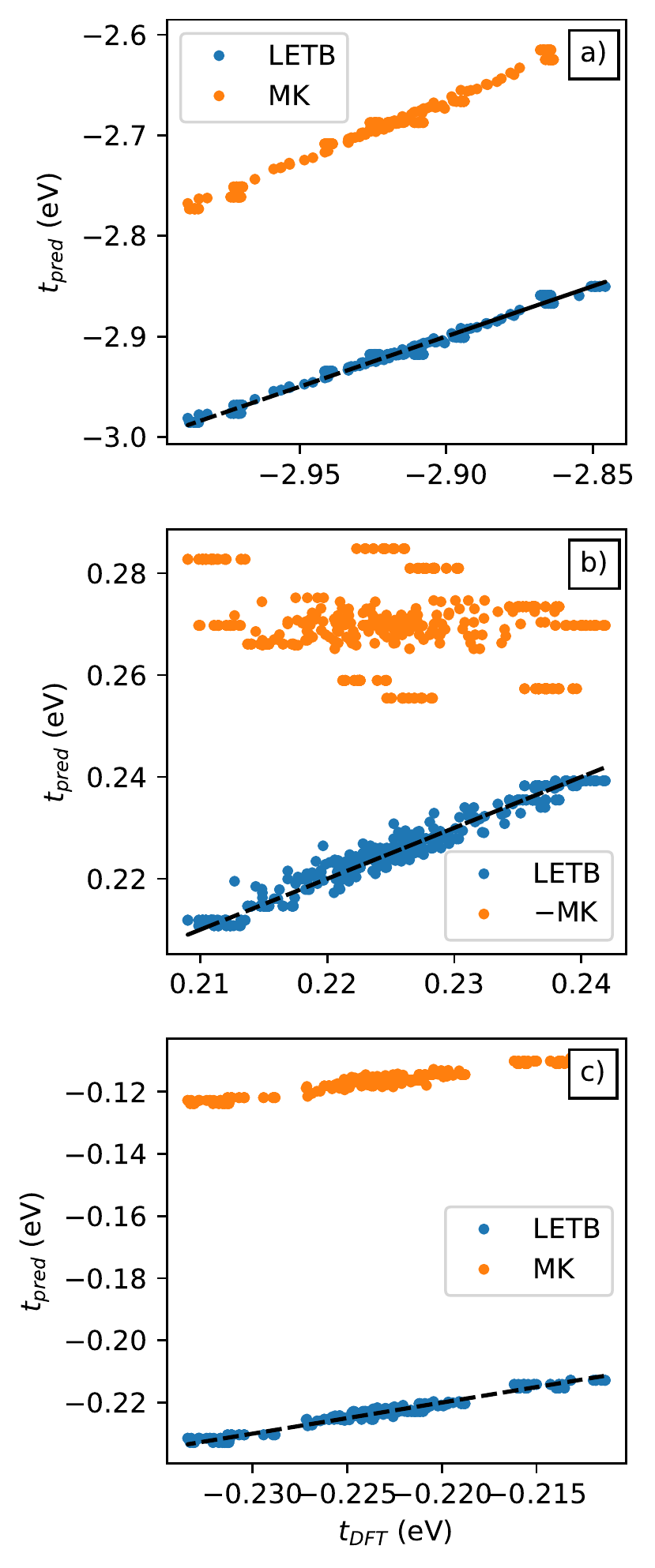}
\caption{Predicted versus computed DFT intra-layer hoppings for a) nearest neighbor, b) next-nearest neighbor, c) third nearest neighbor.
LETB refers to this work, and MK is using the MK parameterization. 
The black line indicates perfect agreement with the computed DFT values.
Note for next-nearest neighbor hopping we present the MK hopping with a sign flip, as the model does not predict negative values.
}
\label{fig:fit_graphene}
\end{figure}

The nearest neighbor hopping is well approximated by a linear function in the distance $a$ between the atoms, as shown in Fig~\ref{fig:fit_graphene}a.
We find excellent agreement to the DFT data with our fit yielding an $R^2$ of 0.98 and root mean square error (RMSE) of 4.22 meV.
The MK parameterization, while following a similar trend to the LETB, consistently underestimates the magnitude of the hoppings by 10\%.

In contrast to the case for nearest neighbors, a parameterization only using interatomic distance fails dramatically for the second and third nearest neighbor hoppings, indicating the need for local environment effects.
As such, the effect of the local environment is included by expanding the pool of potential descriptors to include the descriptors shown in the middle and right panels of Fig~\ref{fig:descriptors}.
Descriptor selection is then carried out using the least absolute shrinkage and selection operator (LASSO)\cite{10.2307/2346178, lasso2} to determine a minimal set of local environment descriptors required to describe variations in intra-layer hoppings.
After descriptor selection is carried out, the linear models 
are fitted using cross validated (CV)\cite{doi:10.1080/00401706.1974.10489157} ordinary least squares regression, with 5 folds.
A summary of the final regressed models with cross validated parameter means and uncertainties are presented in Table~\ref{table:parameters}.

For the second nearest neighbor hopping, LASSO selects the $h_1$ and $h_2$ descriptors before the intersite distance $b$, and discards all other descriptors.
The descriptors $h_1$ and $h_2$ parameterize the shape of the hexagon,  indicating that the hopping is mediated by the shape of the ring itself, rather than direct hopping. 
We find excellent agreement between the three descriptor model and the DFT data, yielding an $R^2$ of 0.92 and RMSE of 1.44 meV as shown in Figure~\ref{fig:fit_graphene}b).
This can be contrasted with the simpler MK parameterization, which predicts a negative hopping parameter and an incorrect dependence on geometry as it does not account for the $h_1, \ h_2$ descriptors, which are the most important for the second nearest hopping.

The third nearest neighbor model in Fig~\ref{fig:fit_graphene}c) is the most complex, involving the size of the entire hexagonal environment around the hopping centers.
The selected descriptors for the LETB are $h, l$ and $c$, with a fit quality of $R^2$ = 0.96 and RMSE of 1.03 meV. 
LASSO selects the distance between the atoms $c$ before the local environment variables $h, l$.
Similar to the first nearest neighbor, the MK parameterization consistently underestimates the hopping magnitude for the third nearest neighbor hopping.
However, in this case the quantitative error is much larger, with a nearly 50\% error between the MK and DFT values.

\subsection{Inter-layer hoppings}
For the inter-layer hoppings,  between a pair of atoms $i, j$ arranged such that one atom resides in either layer of the bilayer, we use the parameterization proposed by Fang and Kaxiras\cite{PhysRevB.93.235153}.
The Fang and Kaxiras model (FK model) takes the following form
\begin{equation}
\begin{split}
t^{FK}_{ij}(\{\vec{R}_{ij}\}) = t^{FK}(\vec{R}_i - \vec{R}_j, \theta_{u, ij}, \theta_{d, ij}) \\ \\
t^{FK}(\vec{r}, \theta_u, \theta_d) = V_0(r_{xy}/a) + \\
V_3(r_{xy}/a)(\cos(3\theta_u) + \cos(3\theta_d)) + \\
V_6(r_{xy}/a)(\cos(6\theta_u) + \cos(6\theta_d))\\
\\
V_0(r) = \lambda_0 e^{-\xi_0 r^2}\cos(\kappa_0r) \\
V_3(r) = \lambda_3r^2 e^{-\xi_3 (r - x_3)^2} \\
V_6(r) = \lambda_6 e^{-\xi_6(r - x_6)^2}\sin(\kappa_6 r).
\end{split}
\label{eq:fang}
\end{equation}
Here the angles $\theta_{u, ij},\  \theta_{d, ij}$ account for the local environment effects between two atoms $i, j$.
The first angle $\theta_u$ indicates the orientation of the nearest neighbor triangle of the \textit{upper} sheet atom relative to the displacement vector $\vec{R}_i - \vec{R}_j$, and $\theta_d$ the orientation of the same triangle of the \textit{lower} sheet atom.
The constant $a = a_0/\sqrt{3} = 1.549$ Bohr, and all other constants in the expression are regression parameters.

We use a non-linear least squares algorithm to fit the parameters in Eq~\ref{eq:fang} to our DFT data.
As with the intra-layer parameters, we use a 5-fold CV to simultaneously fit parameters and assess the error bars for the parameters.
The summary of the fit parameters and their CV error bars are presented in Table~\ref{table:parameters}.
Our fit parameters agree with the original FK model within an order of magnitude and that all parameter signs are consistent.

\begin{figure}
\centering
\includegraphics{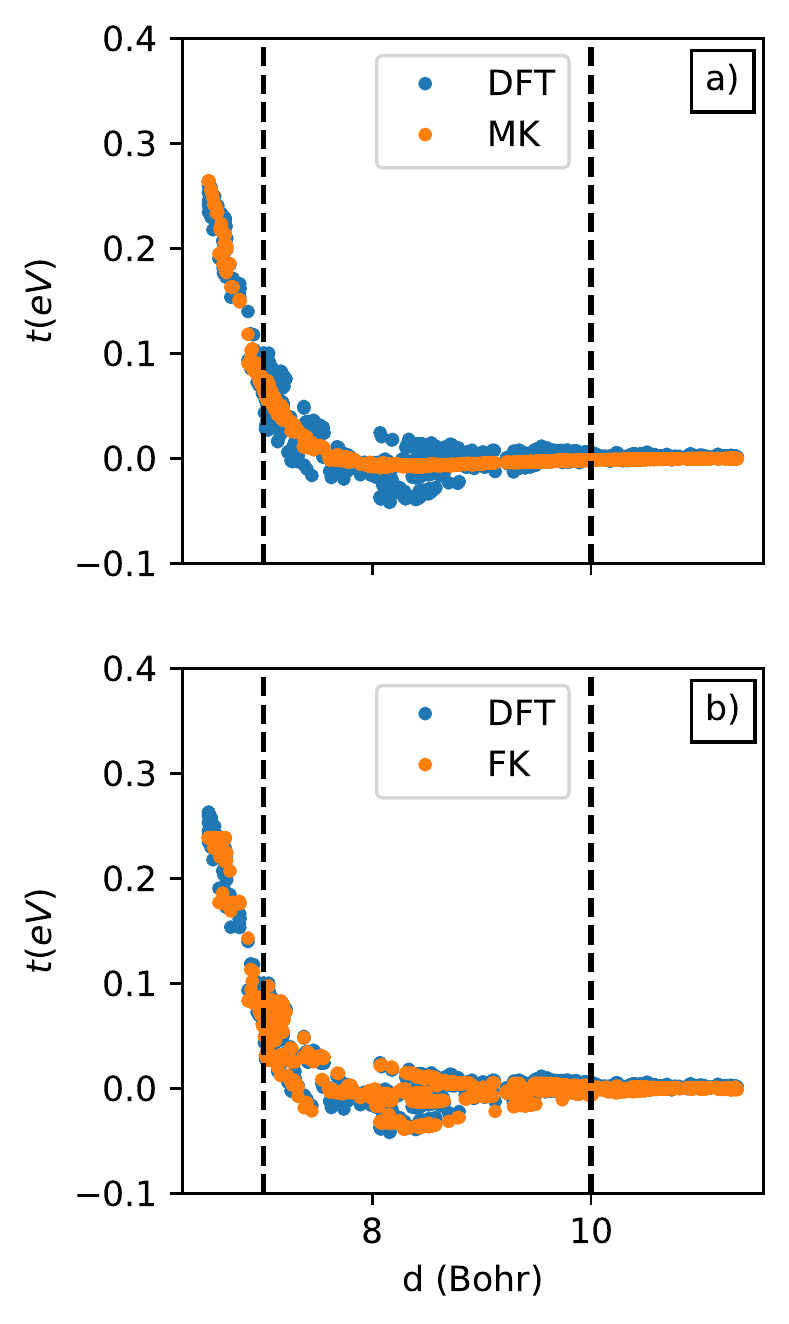}
\caption{Plot of the predicted inter-layer hoppings using the a) MK and b) FK models compared to DFT, as a function of atomic pairwise distance.
Guidelines at 7 Bohr and 10 Bohr highlight the intermediate region 7 Bohr $< d < 10$ Bohr where the MK model fails to describe much of the variation while the FK model performs well. }
\label{fig:t-inter}
\end{figure}

The results of the inter-layer fitting are shown in Fig~\ref{fig:t-inter}.
Looking at the MK parameterization first, we find that the model performs well in two extreme regions --- $d < 7$ Bohr and $d > 10 $ Bohr --- but fails to describe most of the variation in the region in between, near 8 Bohr.
The FK parameterization ameliorates this issue and is able to describe the variation in the DFT data consistently across all bond lengths.
Our results indicate that local environmental effects are required to describe inter-layer hoppings in intermediate bond lengths near 8 Bohr.

\subsection{Combined model}
Hoppings between pairs of atoms $i, j$ in a given twisted configuration are computed using the LETB in a two step process.
First, the hopping between atoms $i, j$ is classified as inter-layer or intra-layer using the pair-wise distance vector $\vec{d} = \vec{R}_i - \vec{R}_j$.
Next, the hopping value is computed using the positions of nearby atoms in the twisted configuration.
Details of the classification method and hopping computation follow.

The hopping is classified as intra-layer if the $z$ projection of the distance $|d_z| <$ 1 Bohr, and inter-layer otherwise.
Intra-layer hoppings are further stratified via the $xy$ distance $d_{xy} = \sqrt{d_x^2 + d_y^2}$ into first, second, and third nearest neighbor classifications if $d_{xy}$ is within $\pm$5\% of the equilibrium distances of $a_0, \sqrt{3}a_0, 2a_0$.
Intra-layer hoppings with $d_{xy} > 1.05 \times 2a_0$ are collected in a single ``out-of-bounds'' category.

Once classified, the hopping is computed using the fit functions in Table~\ref{table:parameters}.
For in-bound classification --- inter-layer hopping and intra-layer hopping up to third nearest neighbor --- all atoms within $2a_0$ distance of atoms $i$ and $j$ are first collected. 
The collection of atoms alongside $i, j$ are then used to compute the appropriate local environment descriptors and subsequent hopping value between $i, j$.
A hopping value of zero is returned for out-of-bound classification.

\subsection{Model validation}
\begin{figure}
\centering
\includegraphics{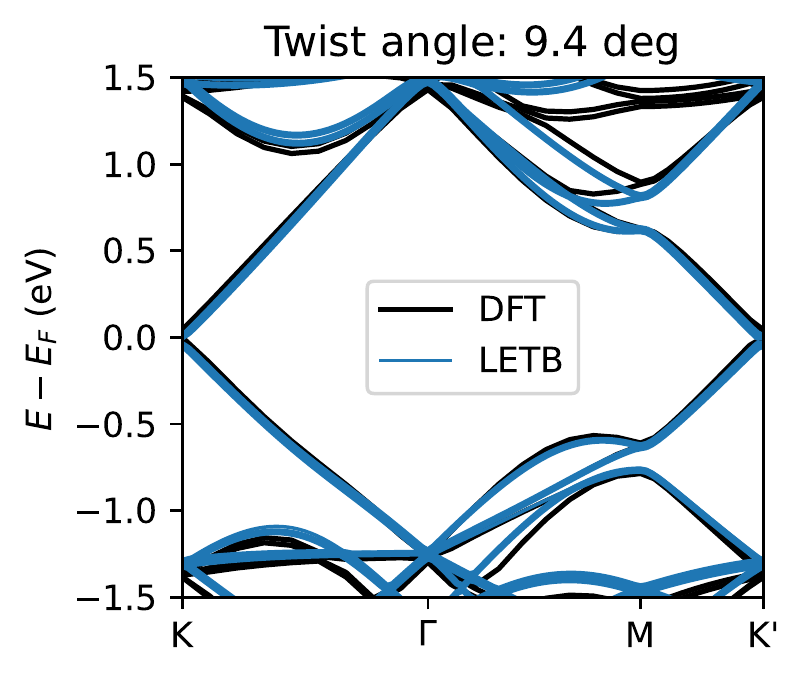}
\includegraphics{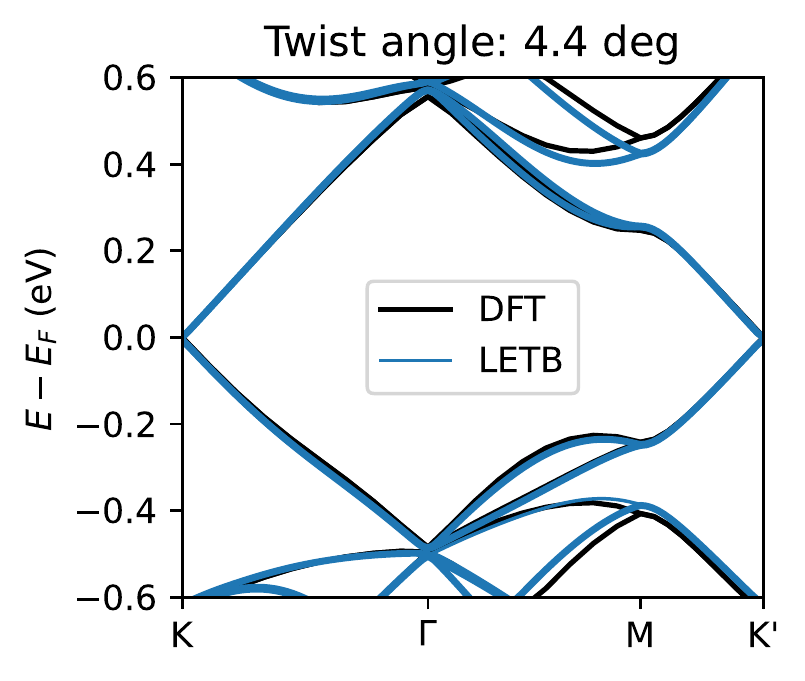}
\caption{Comparison of LETB and MK band structure versus computed DFT band structures for 9.4 and 4.4 degree twist angles.
Energy is relative to the Fermi level, and the standard $K \rightarrow \Gamma \rightarrow M \rightarrow K^\prime$ path through the Moire supercell is shown.}
\label{fig:model-dft}
\end{figure}

As a final check of model validity, in Figure~\ref{fig:model-dft} we compare the DFT band structures for TBLG at 9.4$^\circ$ and 4.4$^\circ$ twists against the band structures computed using LETB.
We find excellent agreement between the DFT and LETB at all k points and at both twists for the four $\pi$ bands near the Fermi surface.
As such, the use of primitive cell bilayer configurations for training paired with a local TB approximation generalizes well to twisted bilayer.

\section{Effects of accurate TB model on the electronic structure of TBLG}
\begin{figure}
\centering
\includegraphics{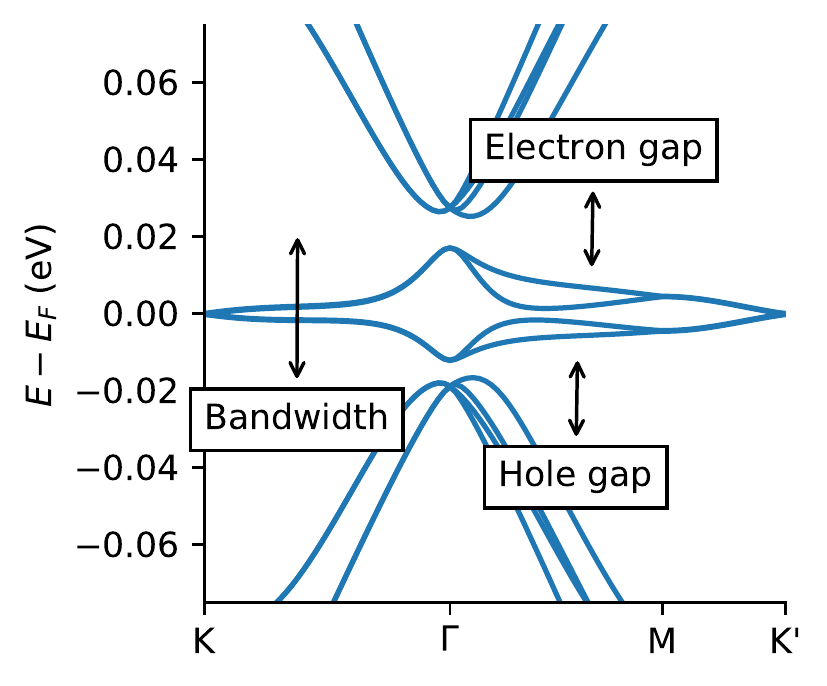}
\caption{The band structure for a rigid 1.05$^\circ$ twist using the LETB model, with important quantities labeled.}
\label{fig:schematic}
\end{figure}

To study the effects of the LETB on our understanding of TBLG,  we computed band structures for various small twist angles using LETB and the MK model with rigid and fully relaxed twisted geometries.
We consider twist angles of 2, 1.47, 1.16, 1.08, 1.05 and 0.99$^\circ$.
Rigid geometries consist of two sheets of unrelaxed graphene sheets twisted relative to each other with a commensurate twist angle at equilibrium AB separation.
The fully relaxed geometries begin with the rigid geometries, and the atomic positions are determined by atomistic molecular static simulations using classical potentials.
Details of the geometry optimization technique and comparison of computed band structures follow. 
A summary of the computed quantities is shown in Figure~\ref{fig:schematic}.

\subsection{Relaxed geometries}
We perform molecular static simulations using the Large-Scale Atomic Molecular Massively Parallel Simulator package (LAMMPS)\cite{PLIMPTON19951} to obtain fully optimized geometries. 
Although a low twist angle TBLG unit cell contains many atoms, it is computationally feasible to optimize the system using classical potentials.
We employ a hybrid of reactive empirical bond order (REBO) potential\cite{brenner_second-generation_2002} for the intra-layer interaction and Kolmogorov-Crespi (KC) potential\cite{kolmogorov_registry-dependent_2005} for the inter-layer interactions. 
Atomic coordinates are relaxed by the centroid-gradient (CG) energy minimization scheme\cite{sheppard_optimization_2008} with a small stopping tolerance of 10$^{-11}$ eV.

\subsection{Bandwidths}
\begin{figure}
\centering
\includegraphics{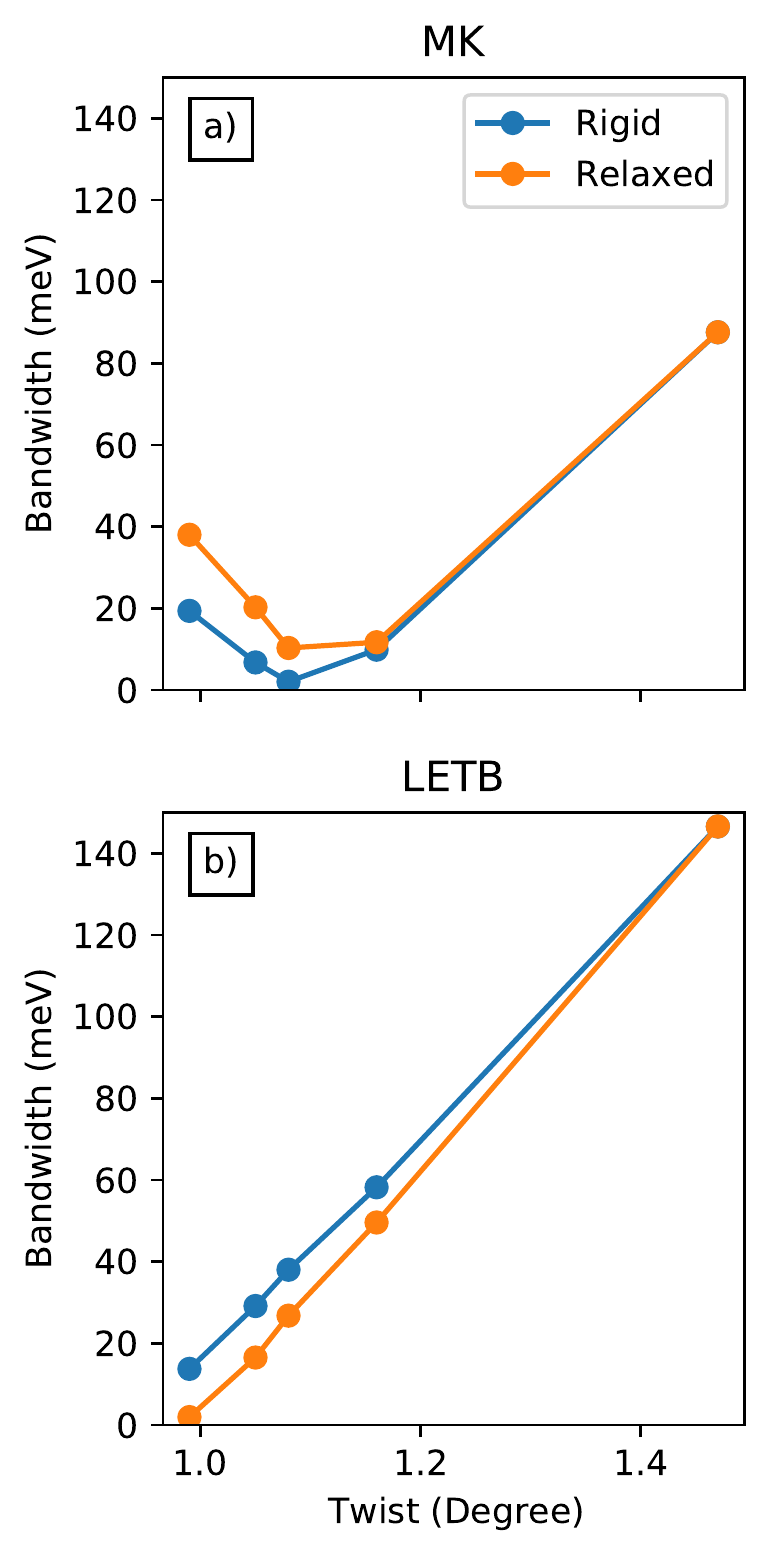}
\caption{Computed flat band bandwidths for small twist angles near the magic twist angle using a) MK and b) LETB model parameterizations.
We consider two different sets of lattice geometries: rigid twisted geometry and a fully relaxed, molecular dynamics geometry. 
}
\label{fig:bw}
\end{figure}

We begin by studying band flattening as described by the LETB and MK models.
In Fig~\ref{fig:bw} we present the bandwidths of the flat bands for the LETB and MK model with rigid and relaxed geometries.
Both models for both geometries yield flat bands with bandwidth below 600 meV for twist angles below 1.2$^\circ$.
With both geometries, the MK parameterization achieves its first inflection point at 1.08 degree twist, whereas the LETB continues the downward trend through 0.99 degree twist. 
It is unclear where the first inflection point for the LETB should be, but it clearly occurs below 0.99$^\circ$ for both geometries.
Defining the first magic twist angle as the first inflection point of the bandwidth with respect to twist, the LETB yields a first magic twist angle at least 10\% smaller than the MK parameterization.

\subsection{Fragile topology}
Fragile topology in TBLG is characterized by the inability to create symmetric, exponentially localized Wannier orbitals with only the minimal set of nearly flat bands in TBLG\cite{PhysRevB.99.195455,  PhysRevLett.123.036401}.
The existence of fragile topology is predicated on the existence of a spectral gap between the minimal set of active bands and the higher (electron) and lower (hole) lying states. We therefore study the electron and hole band gaps (indicated schematically in Fig~\ref{fig:schematic}) as a necessary, but not sufficient, requirement for any non-trivial topology in the flat bands.
We further study the orbital character across the first Brillouin zone, which gives us direct insight into the emergence of Wannier obstruction, and hence the emergence of fragile topology.

\begin{figure}
\centering
\includegraphics{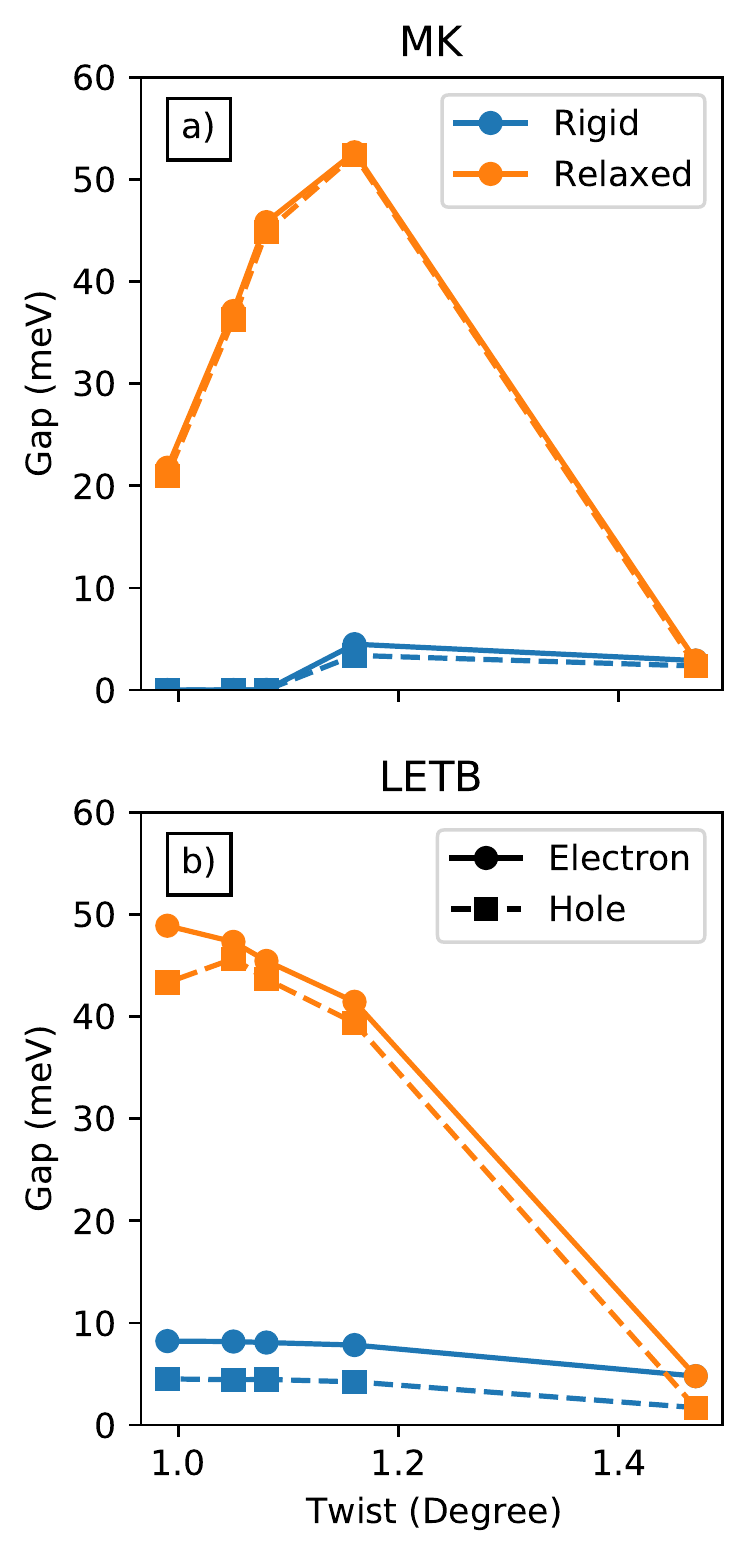}
\caption{Computed electron and hole band gaps for small twist angles near the magic twist angle using a) MK and b) LETB model parameterizations.
We consider two different sets of lattice geometries: rigid twisted geometry and a fully relaxed, molecular dynamics geometry. 
}
\label{fig:gaps}
\end{figure}

In Fig~\ref{fig:gaps} we present the band gaps between the flat and dispersive bands for the LETB and MK model with rigid and relaxed geometries. 
Band gaps were computed for both single-electron and single-hole excitations.
Beginning with the rigid geometry, we note that the LETB has a finite band gap for all twist angles.
The MK parameterization, however, yields zero band gap for twists of 0.99, 1.05 and 1.08$^\circ$, coinciding with the first inflection point observed in Fig~\ref{fig:bw}.
With the relaxed MD geometries, both LETB and MK follow qualitatively similar trends, with band gaps nearly 10 times that of the rigid geometries across all twist angles.
Our results indicate that fragile topology cannot emerge in the MK model for rigid geometries in the vicinity of the first magic twist angle, but that fragile topology may emerge for the same twists when using the LETB model.

\begin{figure}
\includegraphics[scale=0.24]{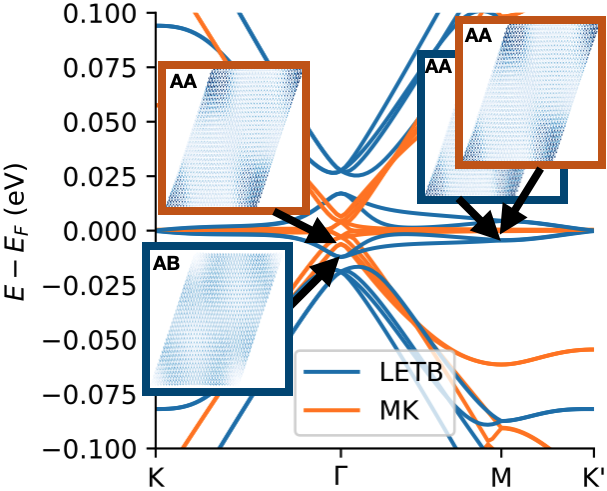}
\caption{Computed band structure and electron density for $\Gamma, M$ points for a rigid 1.05$^\circ$ twisted geometry using the LETB and MK models.
Densities shown are for the bands between the Fermi level with regions of higher density having darker color.
The coordination pattern of the density, AA or AB, is labeled for each density.
}
\label{fig:topology} 
\end{figure}

To directly probe Wannier obstruction in the gapped flat bands, we look at the orbital density across the first Brillouin zone with a rigid 1.05$^\circ$ twisted geometry.
As show in Figure~\ref{fig:topology}, the MK model does not exhibit Wannier obstruction as the electron density in the flat bands localizes to AA character across the Brillouin zone.
However, the LETB model does exhibit Wannier obstruction as the orbital character near $\Gamma$ abruptly changes to AB character, indicating the emergence of a fragile topology in the flat bands.
We also note that Wannier obstruction does not exist in the MK model for  twists of 0.99 and 1.08$^\circ$ when using rigid geometries, coinciding with the band gap closing in Fig~\ref{fig:gaps}.

The computed band gaps and orbitals in the LETB indicate that geometry relaxation may not be required for isolated bands and fragile topology of the flat bands in the LETB model near the first magic twist angle, in contrast to the picture from the MK parameterization.
In the LETB model the AB character at $\Gamma$ is observed even with rigid geometries, preventing the localization of Wannier orbitals within the flat bands.
This localization pattern associated with Wannier obstruction has also been previously observed and linked to the emergence of fragile topology in extended tight binding models using relaxed geometries\cite{PhysRevResearch.1.033072}.
While this is not definitive evidence that LETB houses fragile topology, the emergence of a finite band gap and Wannier obstruction strongly support the emergence of fragile topology with rigid twists near the first magic twist angle.
A conclusive proof would require computing the appropriate topological invariant which distinguishes the trivial and fragile topological flat bands, however there is no consensus currently on the appropriate invariant for TBLG.

\section{Conclusion}
We developed a local environment position-dependent tight-binding model (LETB) that faithfully reproduces density functional theory bands on 72 random configurations of bilayer graphene. 
Transferability was tested versus DFT band structures at 4.4$^\circ$ and 9.4$^\circ$, resulting in very small errors in the computed band structure.
We found that for second and third neighbors, the tight binding parameters are not well-described by the distance between sites, but instead are best described using many-atom descriptors, as encoded in the LETB.
The model is implemented in a software package available online\cite{letb} and integrated with the \texttt{pythtb} package.\cite{Yusufaly2013TightBindingFI}. 

Compared to a simpler parameterization by Moon and Koshino,\cite{PhysRevB.85.195458}, we find isolated flat bands with Wannier obstruction (known as \textit{fragile topology}) without necessarily requiring structural distortions. 
A local minimum in the bandwidth attained at an angle slightly lower than 1.05$^\circ$, reaching zero at around 0.99$^\circ$,  which we identify with the first magic angle. 
The variation in the minimum bandwidth between tight-binding models is perhaps surprising, since one might assume that it is mainly dependent on low-energy physics.
We find that the isolation of the flat bands is quantitatively similar between the two models, but that the MK model likely incorrectly closes the electron and hole gaps for undistorted (rigid) graphene, which qualitatively changes the nature of the states. 
Namely, we found that the MK model cannot house fragile topology near the first magic twist angle with rigid geometries, whereas the emergence of Wannier obstruction in the LETB indicates the presence of fragile topology even with rigid twists.

Our study allows us to make some comments about what parameters affect the bandwidth and electron/hole gaps in twisted bilayer graphene. 
The bandwidth appears to be primarily sensitive to the tight-binding parameterization, with only a small dependence on the distortions of the layers. 
On the other hand, the electron and hole gaps are primarily sensitive to the distortions of the layers, and secondarily dependent on the tight-binding parameterization. 
However, as indicated above, the proposed parameterization can change the qualitative nature of the bands, including the emergence of fragile topology, in the case of rigid layers of graphene. 

Having a highly accurate tight-binding model has partially unraveled the many complex interactions in twisted bilayer graphene. 
We hope that in future studies, interactions can be added to this model to further disentangle the effects of band structure, atomic structure, and electronic interactions in twisted bilayer graphene.

\begin{acknowledgments}
This material is partially based on the work supported by the U.S. Department of Energy, Office of Science, Office of Basic Energy Sciences, Computational Materials Sciences Program, under Award No. DE-SC0020177. TR and EE also acknowledge support from NSF Grant Nos. 1555278 and 1720633  A.H.N. was also supported by the Robert A. Welch Foundation grant C-1818.
\end{acknowledgments}

\appendix*
\section{Detailed checks for DFT and Wannierization}
\begin{figure}
\centering
\includegraphics{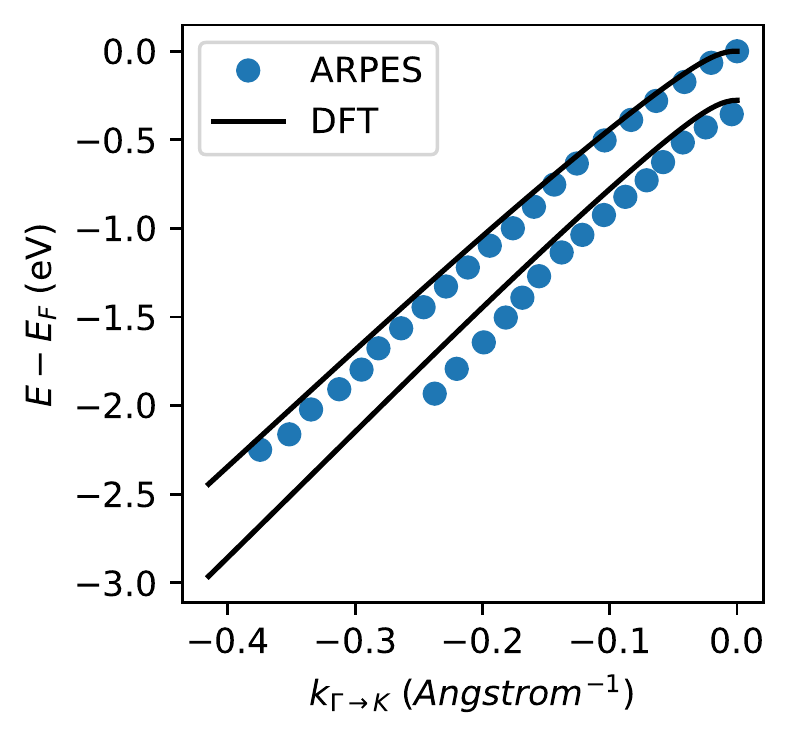}
\caption{Comparison of DFT and ARPES band structures for AB bilayer graphene near the Fermi level. 
The band path goes in a line from the $\Gamma$ to K point in the first Brillouin zone.
ARPES measurements are digitized from Figure 3d of Joucken \textit{et al.}\cite{PhysRevB.99.161406}}
\label{fig:dft-exp}
\end{figure}

To demonstrate the accuracy of the DFT calculations, we present a comparison of the DFT band structure to angle resolved photoemission spectroscopy (ARPES) measurements\cite{PhysRevB.99.161406} in Fig.~\ref{fig:dft-exp} for the eqilibrium AB configuration. 
Near the Fermi level we find excellent agreement between DFT and ARPES, with DFT being within 0.1 eV of the experimentally measured excitations.
The errors between DFT and ARPES increase away from $K$, but do not deviate more than 10\% from the experimental measurements.

\begin{figure}
\centering
\includegraphics{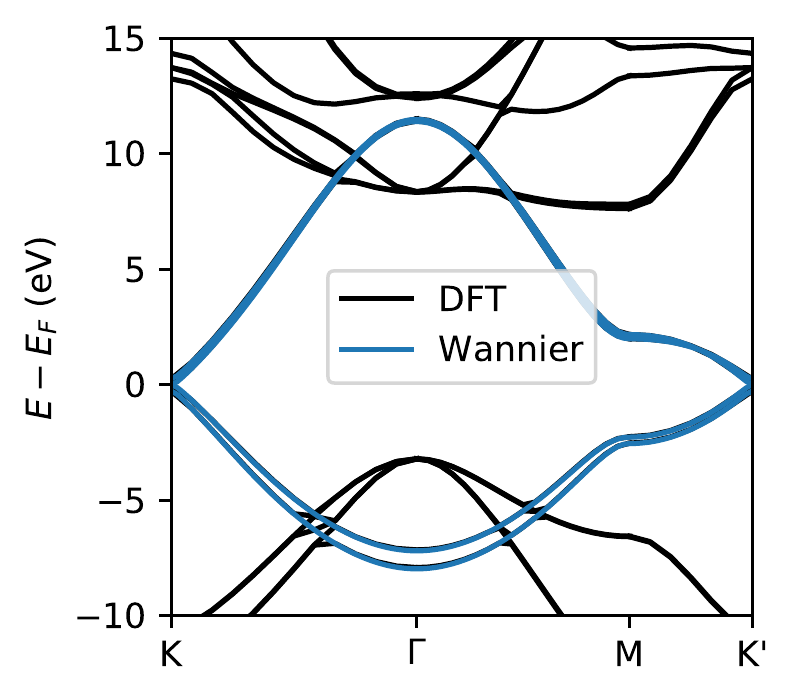}
\caption{Comparison of DFT and Wannierized band structures using the MLWO scheme near the Fermi level for AB bilayer graphene.
Only the $\pi$ bands near the Fermi level are Wannierized.}
\label{fig:wannier-dft}
\end{figure}

We also present a comparison of the Wannierized band structure to DFT is shown in Fig~\ref{fig:wannier-dft} demonstrate the accuracy of the Wannierization procedure.
The figure shows the energy relative to the Fermi level for the DFT and Wannierized bands for the AB configuration.
The Wannierized bands fall exactly on top the DFT bands,  demonstrating the accuracy of both the disentanglement and MLWO schemes.

\bibliography{biblio}
\end{document}